\documentclass[aps,pra, twocolumn,superscriptaddress,floatfix]{revtex4-1}
\usepackage{amsmath,amsfonts,amssymb}
\usepackage{graphicx,color}
\usepackage{epstopdf}
\usepackage[english]{babel}
\usepackage{natbib}
\usepackage{hyperref}
\usepackage[utf8]{inputenc}

\usepackage{letltxmacro}

\LetLtxMacro{\ORIGselectlanguage}{\selectlanguage}
\makeatletter
\DeclareRobustCommand{\selectlanguage}[1]{%
  \@ifundefined{alias@\string#1}
    {\ORIGselectlanguage{#1}}
    {\begingroup\edef\x{\endgroup
       \noexpand\ORIGselectlanguage{\@nameuse{alias@#1}}}\x}%
}
\newcommand{\definelanguagealias}[2]{%
  \@namedef{alias@#1}{#2}%
}
\makeatother

\definelanguagealias{en}{english}
\definelanguagealias{eng}{english}
\definelanguagealias{English}{english}

\let\originalleft\left
\let\originalright\right
\renewcommand{\left}{\mathopen{}\mathclose\bgroup\originalleft}
\renewcommand{\right}{\aftergroup\egroup\originalright}

\newcommand{\ie}{\emph{i.e.}}

\newcommand{\braket}[2]{\left< #1,#2\right>}
\providecommand{\abs}[1]{\left|#1\right|}

\begin{document}
\title{From Many-Particle Interference to Correlation Spectroscopy}

\author{Mattia Walschaers} 
\email{mattia@itf.fys.kuleuven.be}
\affiliation{Physikalisches Institut, Albert-Ludwigs Universit\"at Freiburg, Hermann-Herder-Str. 3, D-79104 Freiburg, Germany}
\affiliation{Instituut voor Theoretische Fysica, KU Leuven, Celestijnenlaan 200D, B-3001 Heverlee, Belgium}
\author{Jack Kuipers}
\affiliation{D-BSSE, ETH Z\"urich, Mattenstrasse 26, 4058 Basel, Switzerland}
\author{Andreas Buchleitner}
\email{a.buchleitner@physik.uni-freiburg.de}
\affiliation{Physikalisches Institut, Albert-Ludwigs Universit\"at Freiburg, Hermann-Herder-Str. 3, D-79104 Freiburg, Germany}

\date{\today}

\begin{abstract}
We show how robust statistical features of a many-particle quantum state's two-point correlations after transmission
through a multi-mode random scatterer can be used as a sensitive probe of the injected particles' mutual indistinguishability. 
This generalizes Hong-Ou-Mandel interference 
as a diagnostic tool for many-particle transmission signals across multi-mode random scatterers. Furthermore, we show how, from such statistical features 
of the many-particle interference pattern, information can be deduced on the temporal structure of the many-particle input state, by inspection of the 
many-particle interference with an additional probe particle of tuneable distinguishability.
\end{abstract}

\maketitle

\section{Introduction}
Identical particles are of profound importance in nature: Pauli's exclusion principle for fermions forms the cornerstone of chemistry, whereas bosonic quantum statistics allows us to prepare Bose-Einstein condensates and induces Planck's law of black body radiation. As such, quantum statistics describes fundamental symmetry properties of quantum states of identical
particles, in equilibrium. Yet, it turns out that the quantum {\em dynamics} of identical particles holds additional and non-trivial surprises, due to intricate 
interference phenomena on the level of many-particle transition amplitudes. The simplest manifestation thereof is the by now well-established 
Hong-Ou-Mandel (HOM) interference dip \cite{hong_measurement_1987,shih_new_1988} which is observed when two photons are transmitted through a balanced beam splitter. However, it recently
has been realised that HOM is only the 
tip of the iceberg of a whole zoo of many-particle interference phenomena \cite{campos_quantum-mechanical_1989,belinskii_interference_1992,lim_generalized_2005, beenakker_two-photon_2009,tichy_zero-transmission_2010,cherroret_entanglement_2011,mayer_counting_2011, tichy_four-photon_2011,tichy_entanglement_2011,mayer_many-particle_2012, tichy_many-particle_2012,nisbet-jones_photonic_2013,ra_nonmonotonic_2013,ra_observation_2013,laibacher_physics_2015,tamma_multiboson_2015,walschaers_quantum_2016,sokolovski_symmetry-assisted_2016,urbina_multiparticle_2016, shchesnovich_universality_2016}, with many particles transmitted through many, randomly coupled modes as the 
other (truly complex) extreme, of potential relevance for photonic quantum simulation and/or computation \cite{bentivegna_experimental_2015,broome_photonic_2013,spagnolo_experimental_2014,spring_boson_2013,tillmann_experimental_2013}. 

\begin{figure}
  	\includegraphics[width=0.46\textwidth]{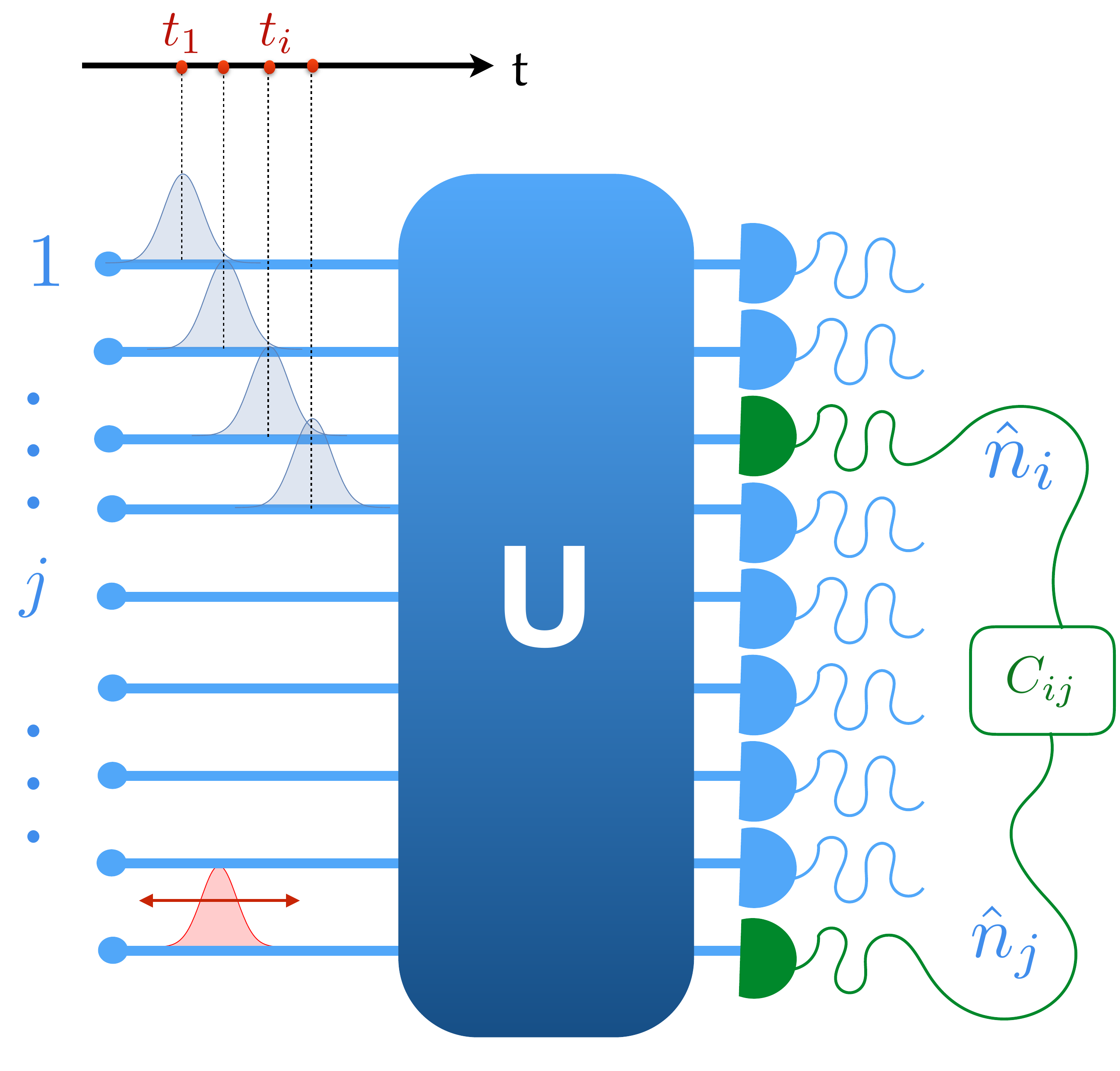}
  \caption{Sketch of the proposed setup. $m$ input modes are connected by a linear optical circuit to $m$ output modes, on each of which a photon counter is mounted. The initial input state to be characterised consists of $n$ (here four -- topmost modes on the left) 
  photons which are described by wave packets. The photons 
  are injected at possibly distinct times $t_j$, which modulates their mutual indistinguishabilities. 
The two-point truncated correlation function $C_{ij}$, eq.~(\ref{eq:CorrFormal}), is the experimental observable to be sampled over all $i\ne j$.
A fifth photon (bottom left) may be injected with controlled arrival time, to probe the temporal structure of the four photon input state, through 
``correlation resonances'' e.g. of the normalised mean $NM$ of the set of $C_{ij}$ (the C-dataset), see eq.~(\ref{eq:spectro}) and Fig.~\ref{fig:Spectro}.
}
   \label{fig:Sketch}
\end{figure}

Hence, many-particle interference defines a new, wide, and rather unexplored area of quantum effects which are indicative, e.g., of the entanglement properties of the many-particle input state \citep{beenakker_two-photon_2009,cherroret_entanglement_2011}, manifest on the semiclassical level \citep{urbina_multiparticle_2016}, and may also be considered as novel resources e.g.
for quantum information processing \cite{bremner_classical_2010, aaronson_computational_2013}. As for all interference phenomena, however, remains the question of their robustness against decoherence effects,
with partial distinguishability \cite{tichy_entanglement_2011,mayer_many-particle_2012, urbina_multiparticle_2016,spagnolo_experimental_2014,shchesnovich_partial_2015,tichy_sampling_2015} of the interfering particles as its arguably most prominent source. Again, HOM and many-particle generalisations thereof 
already provide answers for simple topologies of the coupled modes \cite{belinskii_interference_1992,tichy_four-photon_2011,ra_nonmonotonic_2013,ra_observation_2013}. 

But how does partial distinguishability impact on {\em statistical} quantifiers \cite{mayer_counting_2011,mayer_many-particle_2012,walschaers_statistical_2016,rigovacca_non-classicality_2016} 
of the fine structure of many-particle interference, in cases where the complexity of the dynamics (measured by the number of many-particle amplitudes which are dynamically 
superimposed) prevents a deterministic description, as in principle given, e.g., by evaluation of the full counting statistics? And to which extent can also such statistical quantifiers then be employed as diagnostic tools? 
We here provide the framework for a systematic approach to these questions.

\section{Model}
To do so in a concrete way, we focus on a photonic setup \footnote{This model can easily be adapted to describe free fermions \cite{walschaers_efficient_2016} by considering anticommutation relations for the creation and annihilation operators. To translate our techniques to realistic cold atom scenarios, technical extensions in the description of the scattering compound are required.}. Let us first collect the essential technical tools:
Bosonic Fock space is constructed \cite{bratteli_operator_1997,alicki_field-theoretical_2010} by the vacuum state $\Omega$ acted upon by creation operators of type $a^{\dag}_j(\psi)$. 
The latter creates a photon in the $j$th input mode of the linear optical circuit depicted in Fig.~\ref{fig:Sketch}, with 
the argument $\psi\in\mathcal{H}_{\rm add}$ a state vector 
from 
an auxiliary Hilbert space $\mathcal{H}_{\rm add}$,   
which summarises all additional degrees of freedom of the photon, such as the temporal structure of the incoming wave packets sketched in the figure. 
With the adjoint  
annihilation operators $a_j(\phi)$, 
the associated commutation relations read \cite{alicki_field-theoretical_2010}
\begin{equation}
[a_i(\phi), a^{\dag}_j(\psi)] = \delta_{ij} \braket{\phi}{\psi}.
\end{equation}  
The action of the 
optical circuit in Fig.~\ref{fig:Sketch} is described by an $m \times m$ unitary matrix $U$, with $m$ the number of modes:
\begin{equation}
a^{\dag}_j(\psi) \mapsto \sum_{k=1}^m U_{jk}a^{\dag}_k(\psi).
\end{equation}
Thus, $U$ mixes the different modes, while leaving the additional degrees of freedom untouched. 
An initial state of $n$ photons, prepared in $n$ 
distinct input modes $q_1, \dots, q_n$, 
undergoes the dynamical mapping
\begin{equation}\begin{split}\label{eq:state}
&a^{\dag}_{q_1}(\psi_1)\dots a^{\dag}_{q_n}(\psi_n) \Omega  \mapsto \\&\quad \sum_{k_1, k_2, \dots, k_n=1}^m U_{q_1 k_1}\dots U_{q_n k_n} a^{\dag}_{k_1}(\psi_1)\dots a^{\dag}_{k_n}(\psi_n) \Omega =: \Psi,
\end{split}
\end{equation} 
where we are particularly interested in situations where particle and mode numbers $n$ and $m$, respectively, are significantly larger than in the HOM setting, \ie, $n,m > 2$, but nevertheless far bellow the thermodynamic limit. Hence we explore a widely uncharted parameter regime where one may expect to uncover new physical phenomena, since it is in this parameter range that quantum granularity should be most prominent.

For $n$ and $m$ sufficiently large, and $U$ lacking any prominent symmetry properties, a deterministic evaluation
of $\Psi$ rapidly turns into an intractable problem, and a statistical treatment is needed \cite{walschaers_quantum_2016,walschaers_statistical_2016,walschaers_efficient_2016}. We have shown 
earlier that statistical sampling over the set of 
two-mode truncated correlation functions
\begin{equation}\label{eq:CorrFormal}
C_{ij}:=\langle \hat{n}_i \hat{n}_j \rangle_{\Psi} - \langle \hat{n}_i \rangle_{\Psi} \langle \hat{n}_j \rangle_{\Psi}\, ,
\end{equation}
for all pairs of output modes $i\neq j$,  
allows the detection of robust and characteristic features indicative of the many-particle interferences as induced 
by $U$ for distinguishable and indistinguishable particles \cite{walschaers_statistical_2016}. Here we expand this theory to monitor the continuous 
(quantum-classical, in the sense of quantum statistics) transition from strictly indistinguishable to fully distinguishable
particles, which can be tuned by a continuous degree of freedom accommodated by $\mathcal{H}_{\rm add}$. 
Specifically, we choose this degree of freedom as given by the photon arrival times $t_j$, $j=1,\ldots ,n$ (see Fig.~\ref{fig:Sketch}).

To evaluate (\ref{eq:CorrFormal}) while taking account of the temporal degree of freedom attached to each of the interfering photons, 
we define the single mode number operators on the output by
\begin{equation}
\hat{n}_i := \sum_{k} a^{\dag}_i(\eta_k)a_i(\eta_k),
\end{equation}
where the $\eta_k$ form a basis 
of $\mathcal{H}_{\rm add}$ \footnote{Note that this definition of the number operators implies that the detectors in Fig.~\ref{fig:Sketch} do integrate over the 
photon arrival times.}.
The explicit expression for (\ref{eq:CorrFormal}) then reads
\begin{equation}\begin{split}\label{eq:Corr}
C_{ij} =&  \sum^n_{k\neq l=1} \abs{\braket{\psi_k}{\psi_l}}^2 U_{q_k i}U_{q_l j}U^*_{q_l i}U^*_{q_k j}\\
&-\sum^n_{k=1} U_{q_k i}U_{q_k j}U^*_{q_k i}U^*_{q_k j} \, ,
\end{split}
\end{equation}
with $\psi_k$ the kth photon's wave function in the temporal degree of freedom. The overlap $\abs{\braket{\psi_k}{\psi_l}}^2$, tantamount to the degree
of indistinguishability of the $k$th and $l$th photon (with values between one and zero), gives a tuneable weight to the two-particle interference term 
in (\ref{eq:Corr}), and thus continuously interpolates between the fully indistinguishable ($\abs{\braket{\psi_k}{\psi_l}}^2=1$) and the fully distinguishable
($\abs{\braket{\psi_k}{\psi_l}}^2=0$, for all $k\ne l$) case.

\section{Statistical Certification of Partial Distinguishability}
\subsection{Certification}
Given that the statistics, and, in particular, already the lowest order moments of the {\em C-dataset} \cite{walschaers_statistical_2016} (defined as the sample of all 
$C_{ij}$, $i\ne j$) define unambiguous benchmarks for many-particle interference of (in-)distinguishable particles, (\ref{eq:Corr}) now 
is the fundamental building block to derive analytic expressions for those lowest order moments, for {\em arbitrary} choices of the injected 
photons' mutual indistinguishabilities $\abs{\braket{\psi_k}{\psi_l}}^2$: 
The normalised mean ($NM$) and the coefficient of variation ($CV$) of the C-dataset given by
\begin{align}\label{NM}
NM &:= \frac{m^2}{n} M_1, \\
CV &:= \frac{\sqrt{M_2 - M_1^2}}{M_1}\label{CV}\, ,
\end{align}
with 
\begin{align}\label{eq:M1}
M_1 &:= \frac{2}{m(m-1)} \sum_{i < j = 1}^m C_{ij}, \\
M_2 &:= \frac{2}{m(m-1)} \sum_{i < j = 1}^m (C_{ij})^2\, ,  \label{eq:M2} 
\end{align}
together with the overlap of Gaussian photonic wave packets, centred at $t_j$ with spectral width $\Delta \omega$,
\begin{equation}\label{eq:waveFunc}
\abs{\braket{\psi_k}{\psi_l}}^2 = \exp\Bigg(-\frac{{(\Delta\omega)}^2 (t_k - t_l)^2}{2}\Bigg),
\end{equation}
lead to explicit random matrix theory (RMT) \cite{mehta_random_2004}  predictions, for $U$ given by a random unitary matrix chosen from the Haar measure. From the literature \cite{samuel_un_1980, mello_averages_1990, brouwer_diagrammatic_1996,collins_integration_2006,berkolaiko_combinatorial_2013} we extract the key identity
\begin{equation}\begin{split}\label{eq:average}
\mathbb{E}_U&(U_{a_1,b_1}\dots U_{a_n,b_n}U^*_{\alpha_1,\beta_1}\dots U^*_{\alpha_n,\beta_n})\\ &= \sum_{\sigma,\pi \in S_n}V_m(\sigma^{-1}\pi)\prod^n_{k=1}\delta(a_k-\alpha_{\sigma(k)})\delta(b_k-\beta_{\pi(k)}),
\end{split}
\end{equation}
for the average over $m \times m$ unitary matrices. The functions $V_m(\sigma^{-1}\pi)$ in (\ref{eq:average}) can be obtained via different methods, as shown in \cite{brouwer_diagrammatic_1996, collins_integration_2006}.  The combination of (\ref{eq:average}) with (\ref{eq:Corr}) leads to
\begin{equation}\begin{split}\label{eq:spectro}
&NM \approx \mathbb{E}_{U}\Big(C_{ij}\Big) \frac{m^2}{n}\\
&= -\frac{m}{m+1} \Bigg(1 + \frac{1}{n (m-1)}\sum_{k\neq l =1}^n \exp\bigg(-\frac{{(\Delta\omega)}^2 (t_k - t_l)^2}{2}\bigg)\Bigg)\, ,
\end{split}
\end{equation} 
and 
\begin{equation}\begin{split}\label{eq:random_t}
\overline{NM} &\approx \mathbb{E}_{U}\Big(\overline{C_{ij}}\Big) \frac{m^2}{n}\\
&= -\frac{m}{m+1} \bigg(1 + \frac{n-1}{\sqrt{1+2(\Delta \omega \delta t)^2} (m-1)}\bigg)\, .
\end{split}
\end{equation} 
$NM$ predicts the normalized mean for well-defined injection times $t_k$, while $\overline{NM}$ assumes independently (normal) 
distributed photonic arrival times $t_k$ with zero mean and width $\delta t$, hence implies an additional statistical average over the 
arrival times. In addition, (\ref{eq:average}) allows us to evaluate $CV$ and $\overline{CV}$. We use the results from \cite{brouwer_diagrammatic_1996} in a long but straightforward computation, and obtain
\begin{equation}\label{eq:PDTrainSecondMomentOri}
\begin{split}
\mathbb{E}_{U}({C_{ij}}^2) = &\frac{2 A - 2 B (m-5) + 2 D (2 + 6 m -n + mn) }{(m-1) m^2 (m+1) (m+2) (m+3)}
\\& +  \frac{C (10 + m + m^2) }{(m-1) m^2 (m+1) (m+2) (m+3)}   
  \\& + \frac{(m-2) (1 + 3m)n + 2n^2 + m n^ 2 + 
      m^2n^2)}{(m-1) m^2 (m+1) (m+2) (m+3)},
\end{split}
\end{equation}
with
\begin{align}
&A  = \sum^n_{\substack{k_1,k_2,l_1,l_2 = 1\\ k_1\neq k_2\neq l_1\neq l_2}} \abs{\braket{\psi_{k_1}}{\psi_{l_1}}}^2\abs{\braket{\psi_{k_2}}{\psi_{l_2}}}^2,\label{eq:ASecMom} \\
&B = \sum^n_{\substack{k ,l_1,l_2 = 1\\ k \neq l_1\neq l_2}} \abs{\braket{\psi_{k}}{\psi_{l_1}}}^2\abs{\braket{\psi_{k}}{\psi_{l_2}}}^2, \label{eq:BSecMom}\\
&C = \sum^n_{\substack{k ,l = 1\\ k \neq l}} \abs{\braket{\psi_{k}}{\psi_{l}}}^4, \label{eq:CSecMom} \\
&D = \sum^n_{\substack{k ,l = 1\\ k \neq l}} \abs{\braket{\psi_{k}}{\psi_{l}}}^2. \label{eq:DSecMom}
\end{align} 
Moreover, we find that
\begin{equation}\label{eq:PDTrainSecondMoment}
\begin{split}
\mathbb{E}_{U}(\overline{C_{ij}}^2) = &\frac{2 A' - 2 B' (m-5) + 2 D' (2 + 6 m -n + mn)}{(m-1) m^2 (m+1) (m+2) (m+3)}
  \\& + \frac{ C' (10 + m + m^2) }{(m-1) m^2 (m+1) (m+2) (m+3)}   
  \\& + \frac{(m-2) (1 + 3m)n + 2n^2 + m n^ 2 + 
      m^2n^2)}{(m-1) m^2 (m+1) (m+2) (m+3)},
\end{split}
\end{equation}
with
\begin{align}
&A' = \frac{n(n-1)(n-2)(n-3)}{(1+2(\Delta \omega \delta t)^2)},\label{eq:ASecMomPrime} \\
&B' = \frac{n (n-1) (n-2)}{(1+2(\Delta \omega \delta t)^2)}, \label{eq:BSecMomPrime}\\
&C' = \frac{n(n-1)}{(1+2(\Delta \omega \delta t)^2)} , \label{eq:CSecMomPrime} \\
&D' = \frac{n(n-1)}{\sqrt{1+2(\Delta \omega \delta t)^2}}. \label{eq:DSecMomPrime}
\end{align}
We can combine these outcomes with the results for $NM$ (12) and $\overline{NM}$ (13) to determine
\begin{align}
CV= \frac{\sqrt{\mathbb{E}_{U}({C_{ij}}^2) - \mathbb{E}_{U}({C_{ij}})^2 }}{\mathbb{E}_{U}({C_{ij}}) }, \label{eq:final1} \\ \overline{CV}= \frac{\sqrt{\mathbb{E}_{U}(\overline{C_{ij}}^2) - \mathbb{E}_{U}(\overline{C_{ij}})^2 }}{\mathbb{E}_{U}(\overline{C_{ij}}) } \label{eq:final}.
\end{align}
The latter leads to the RMT prediction for the second panel in Fig.~\ref{fig:Train}.

\subsection{Results for Fluctuating Arrival Times}
Let us first compare the RMT prediction (\ref{eq:random_t}) for $\overline{NM}$ and $\overline{CV}$ to numerically generated results which are obtained by 
direct evaluation of (\ref{eq:M1}) and (\ref{eq:M2}), \ie~as the averages over all possible choices of output modes of a fixed random circuit $U$,
and over Gaussian distributed $t_k$, $k=1,\ldots ,n$, with variable $\delta t$ and fixed $\Delta \omega$. 
\begin{figure}

	\includegraphics[width=0.43\textwidth]{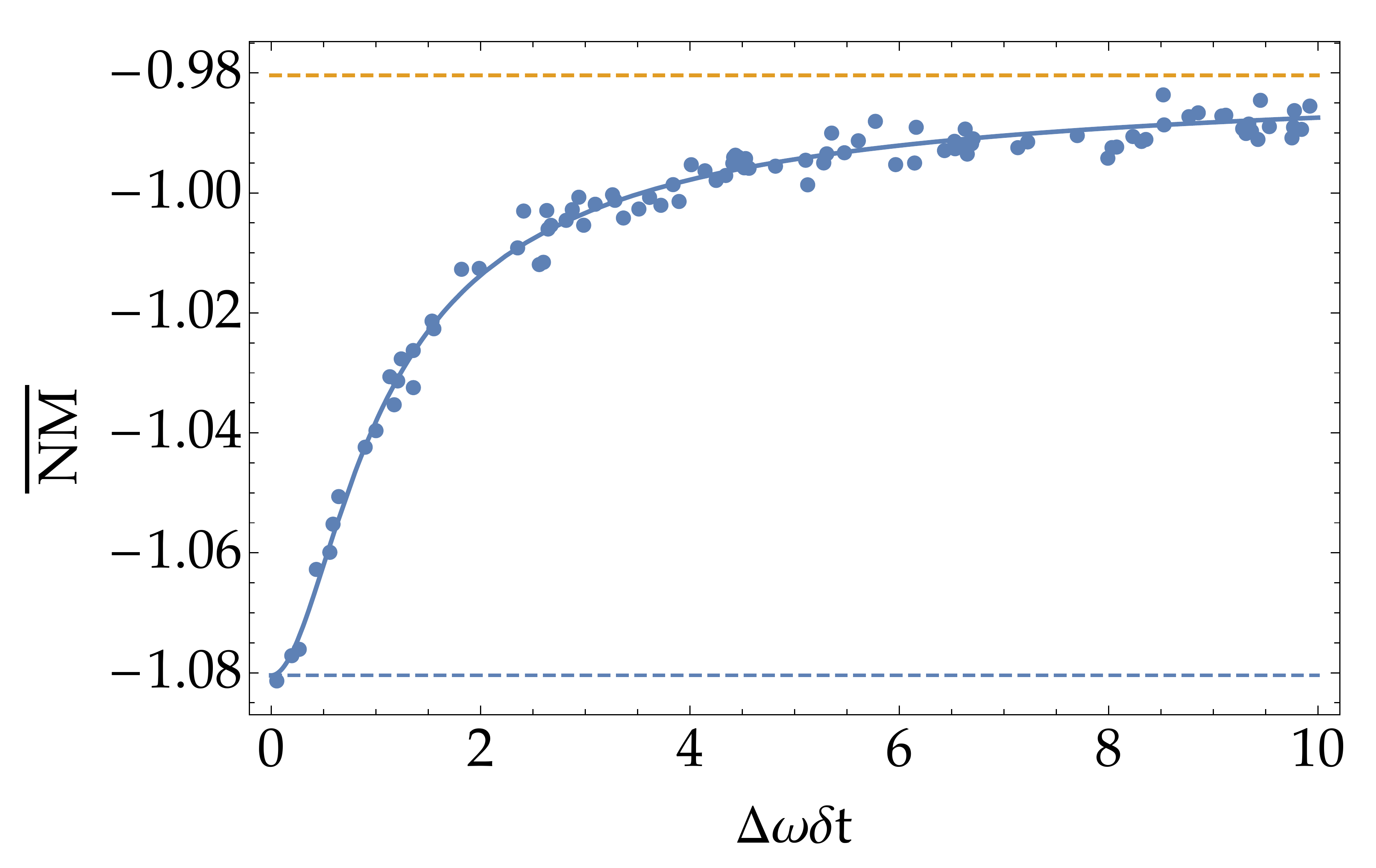}
  	\includegraphics[width=0.43\textwidth]{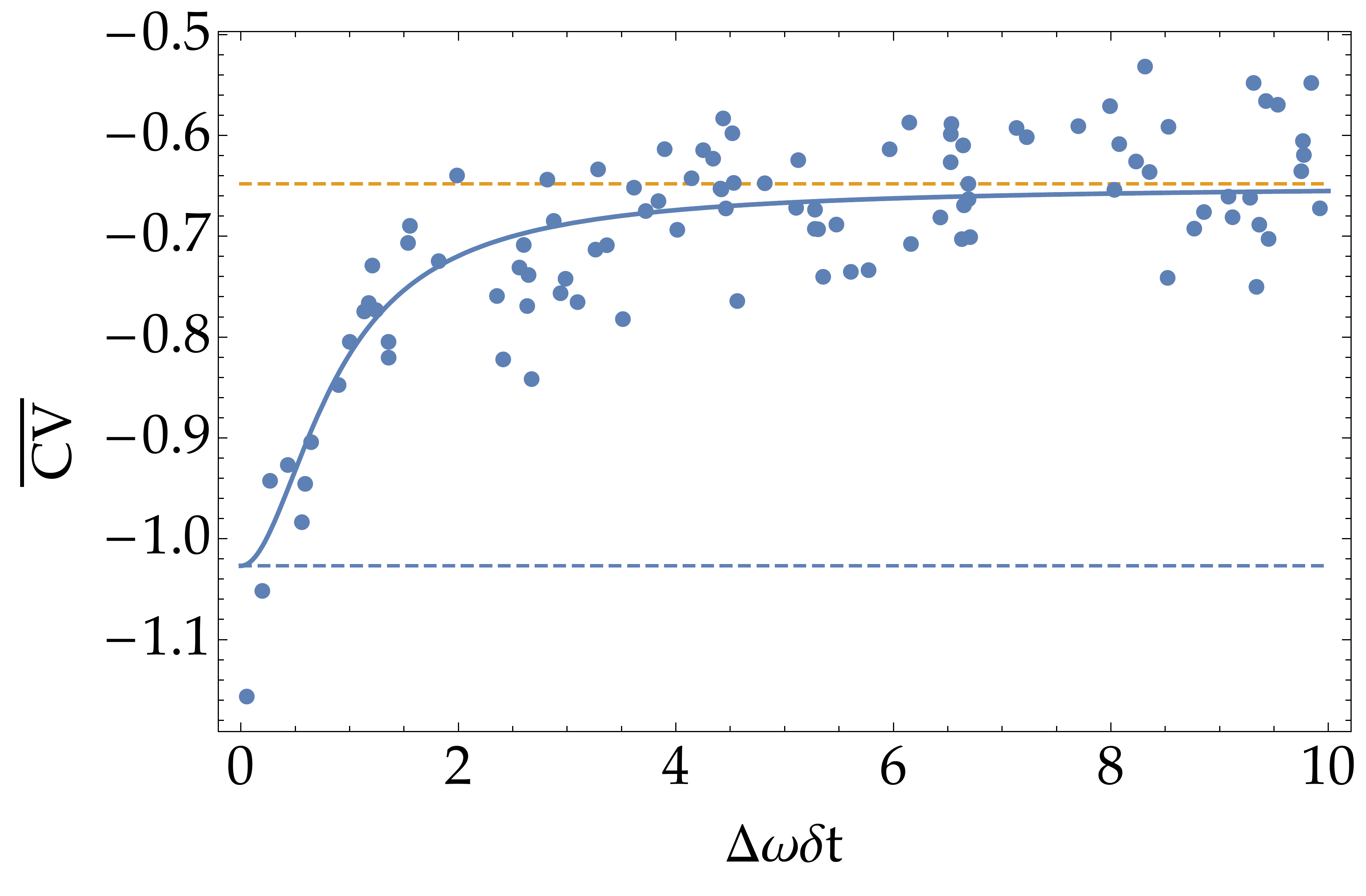}

      \includegraphics[width=0.2\textwidth]{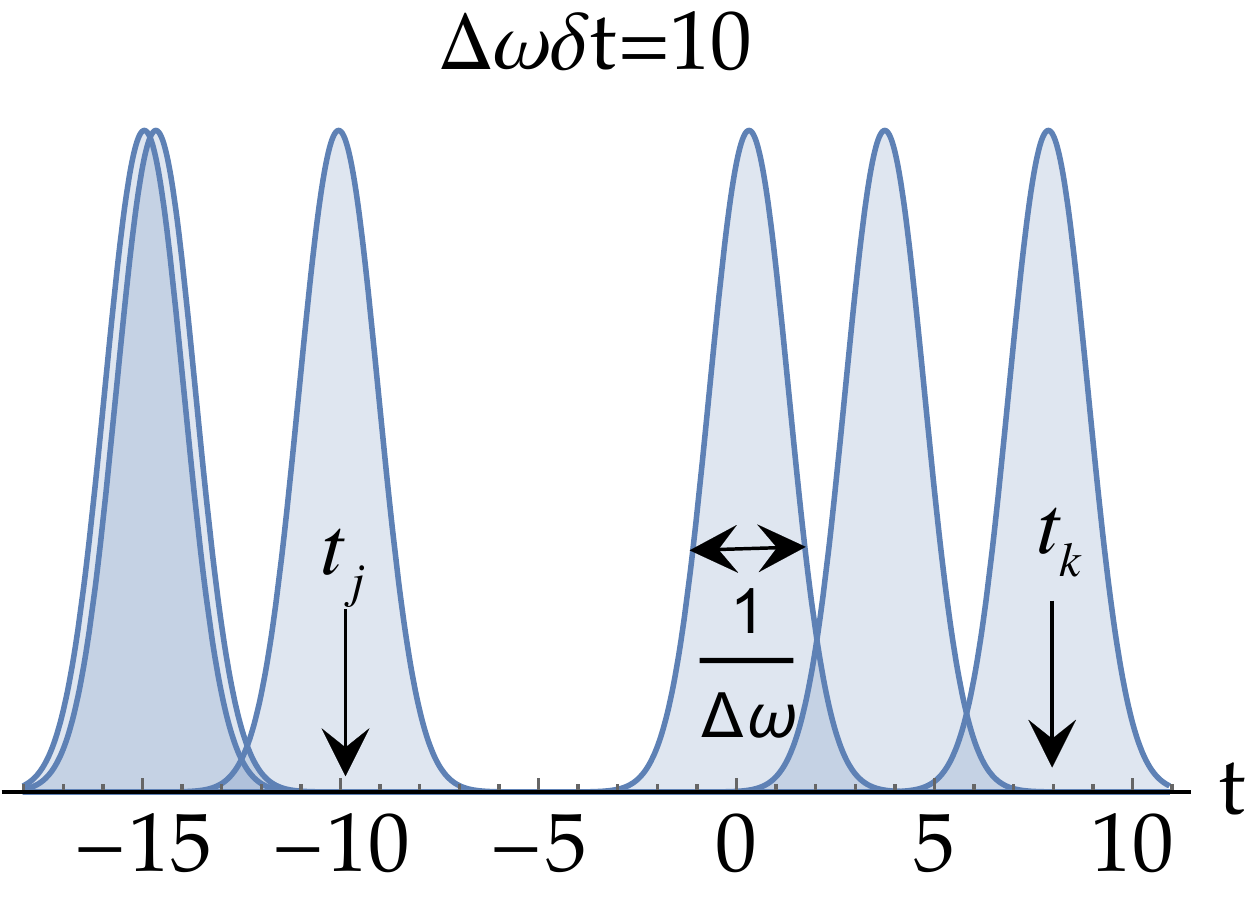}
	\includegraphics[width=0.2\textwidth]{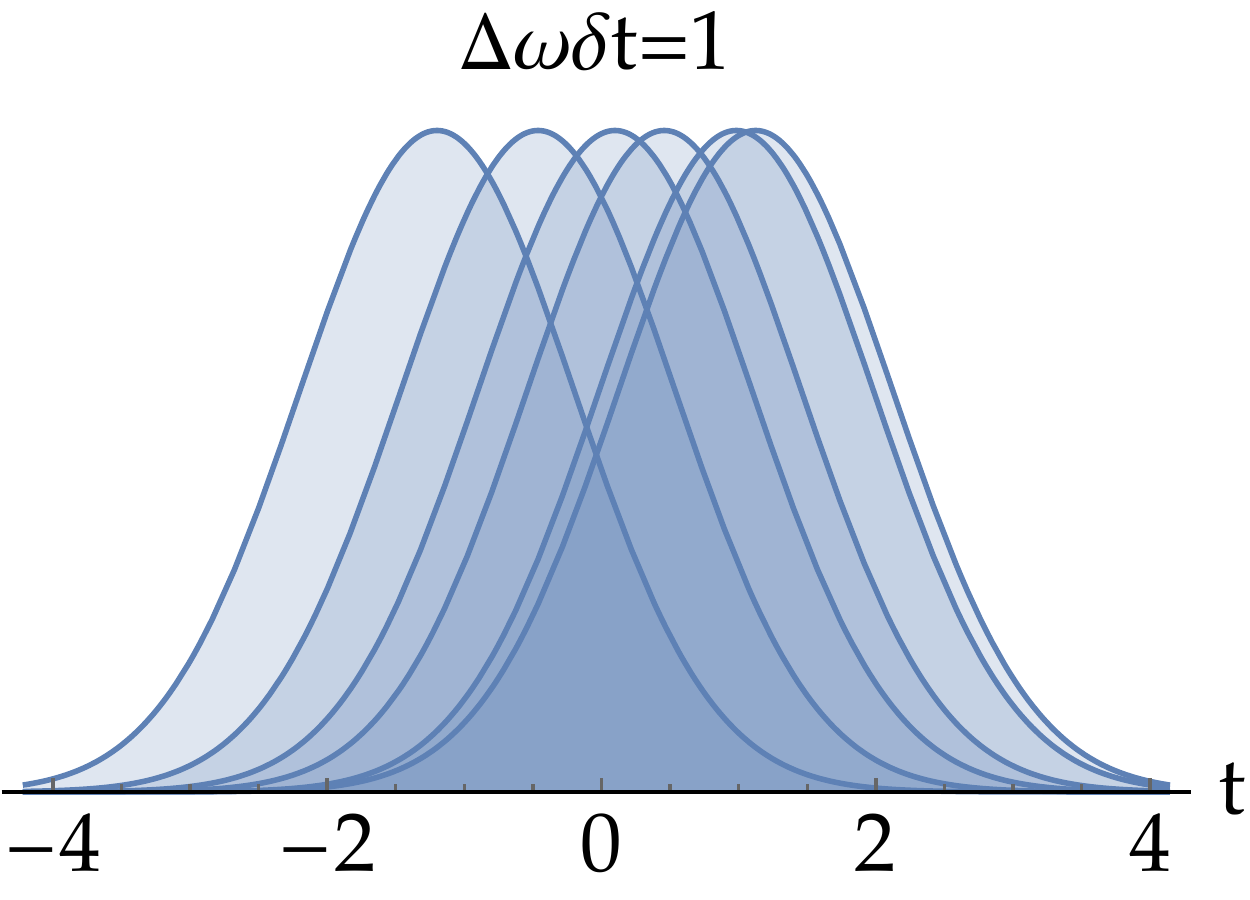}

  \caption{Normalised mean ($NM$, top, eq.~(\ref{NM})) and coefficient of variation ($CV$, middle, eq.~(\ref{CV})) of the C-dataset 
  generated by the transmission of $n=6$ photons across a $m=50$ mode random unitary, as a function of the temporal scatter $\delta t$
  of the photons' arrival times (normally distributed, with zero mean) $t_j$, at given spectral width $\Delta\omega$. The bottom plot sketches
  two typical scenarios of the photons' timing---one where the different wave packets are typically well-resolved (left) and one with a high degree of indistinguishability. Continuous lines indicate the RMT predictions (\ref{eq:random_t}) and (\ref{eq:final}) for $\overline{NV}$ and $\overline{CV}$, while dots are derived from a numerically generated
  C-dataset, with one single, fixed random realisation of $U$, and upon average over $100$ normally distributed arrival times per $t_j$.
  The differences between  the predictions for strictly indistinguishable bosons and for distinguishable particles (horizontal dotted lines) 
  determine the visibilities (\ref{eq:VNM}, \ref{eq:VCV}) of the signals.
  }
   \label{fig:Train}
\end{figure}
Fig.~\ref{fig:Train} shows the {\em statistical} analog of 
the HOM dip, as exhibited by both, $\overline{NM}$ and $\overline{CV}$. Our new, analytical RMT prediction and 
numerical simulation agree very well. Note that residual fluctuations of the numerical result around the RMT prediction, 
more prominent for the coefficient of variation, will be progressively suppressed in the 
thermodynamic limit. 

\subsection{Scaling Behaviour}
\begin{figure}

	\includegraphics[width=0.43\textwidth]{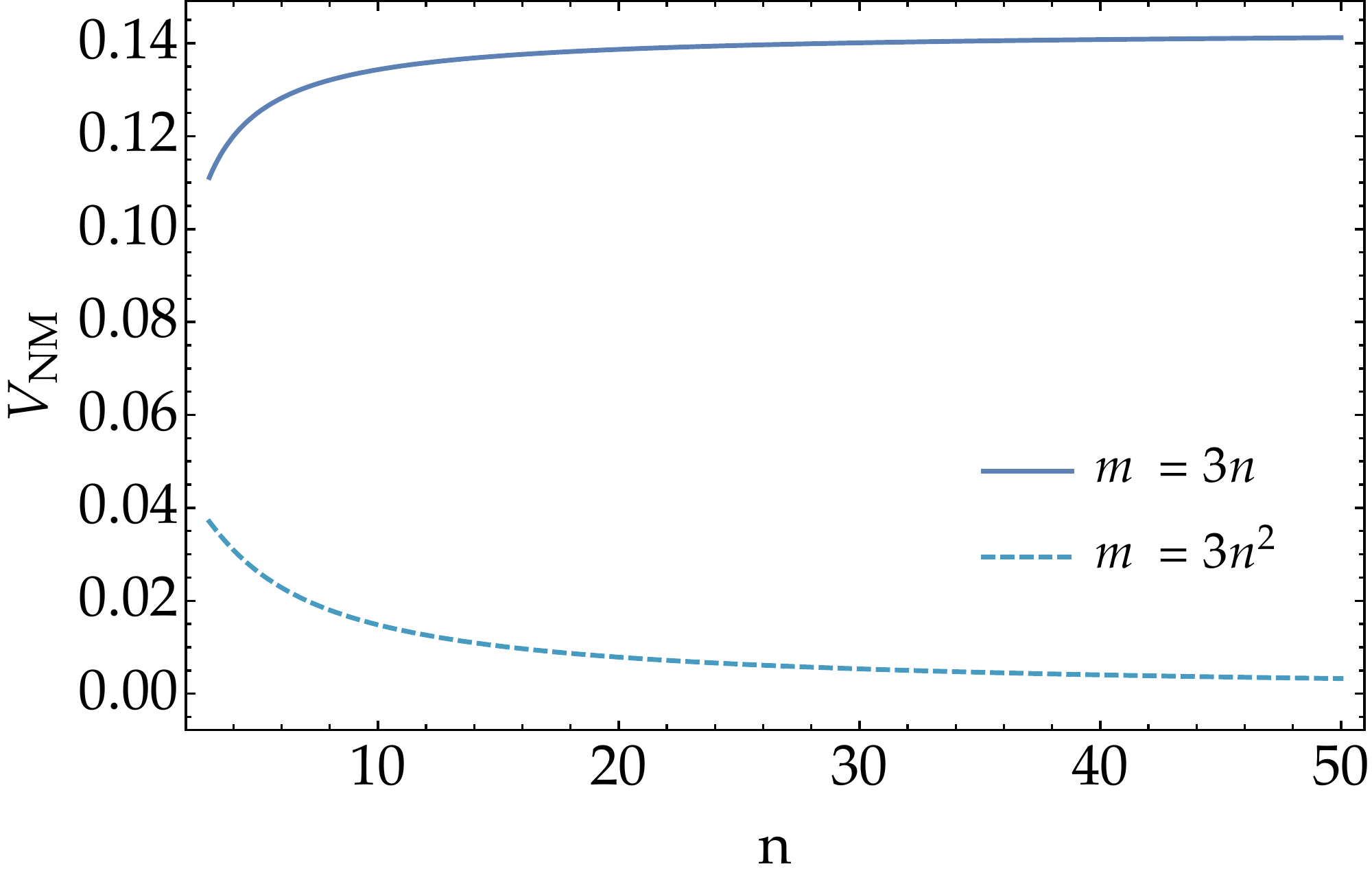}
  	\includegraphics[width=0.43\textwidth]{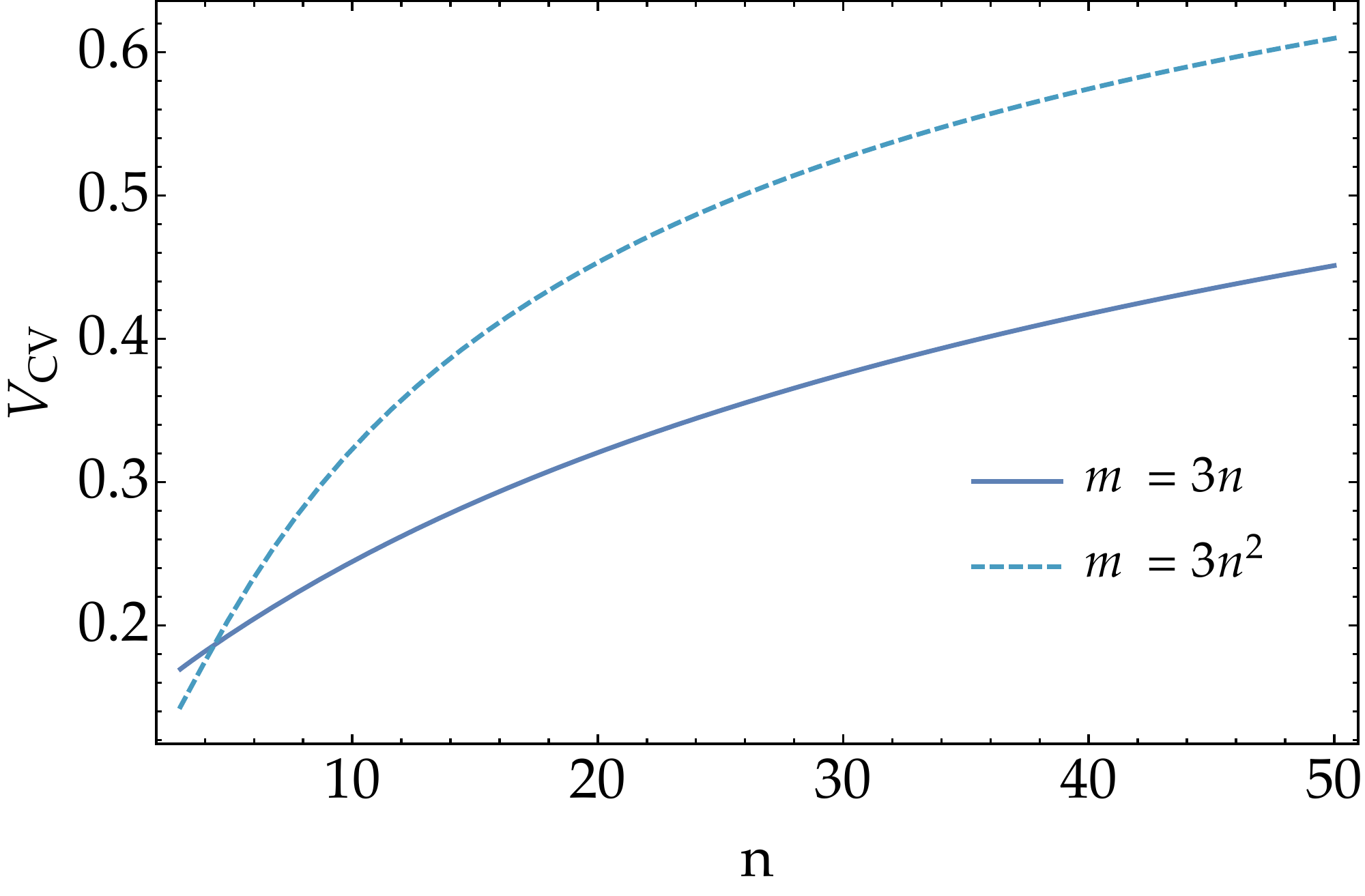}

  \caption{RMT predictions for the visibility of normalised mean ($V_{\rm NM}$, top, eq.~(\ref{eq:VNM})) and coefficient of variation ($V_{\rm CV}$, bottom, eq.~(\ref{eq:VCV})) as obtained from (\ref{eq:random_t}) and (\ref{eq:PDTrainSecondMoment} - \ref{eq:final}). The number $m$ of modes is chosen to scale with the number $n$ of particles as $m = 3 n$ (solid line) and as $m=3n^2$ (dashed line), respectively, where the factor $3$ is chosen arbitrarily.
  }
   \label{fig:Scaling}
\end{figure}

The visibility of the dip is given by the difference between the results for the distinguishable and indistinguishable case:
\begin{align}
&V_{\rm NM} = \abs{\frac{{NM}_{\delta t \rightarrow \infty} - {NM}_{\delta t \rightarrow 0} }{{NM}_{\delta t \rightarrow \infty} + {NM}_{\delta t \rightarrow 0}}},\label{eq:VNM}\\
&V_{\rm CV} = \abs{\frac{{CV}_{\delta t \rightarrow \infty} - {CV}_{\delta t \rightarrow 0} }{{CV}_{\delta t \rightarrow \infty} + {CV}_{\delta t \rightarrow 0}}}.\label{eq:VCV}
\end{align}
The scaling behaviour in $n$ and $m$ is then obtained from (\ref{eq:random_t}) and (\ref{eq:PDTrainSecondMoment} - \ref{eq:final}). In Fig.~\ref{fig:Scaling} we show how these visibilities change as a function of the number $n$ of particles, both in the regime where $m \sim n$, and where $m \sim n^2$ \footnote{Note that boson sampling setups are only proven to be computationally hard for $m > n^2$ \cite{aaronson_computational_2013}.  Thus the scaling $m \sim n^2$ is specifically relevant within this context.}. It becomes clear from the saturation or decrease of $V_{\rm NM}$, and from the monotonous increase of $V_{\rm CV}$ with $n$, that, although the statistical spread in Fig.~\ref{fig:Train} is larger for $\overline{CV}$ than for $\overline{NM}$, the visibility of the distinguishability transition in the former quantity scales more favourably with the system size. This implies that, specifically in the regime of larger $n$ and $m$---and hence where the RMT prediction is more accurate  \cite{walschaers_statistical_2016}---the clearest transition from indistinguishable to distinguishable photons is seen in $\overline{CV}$. 

Therefore, much as in the HOM setting, but now for {\em large} $n$ and $m$, for {\em unknown, random} $U$, and {\em on the level of the lowest order statistical moments of the set of two-point correlation functions read off from the $n$-particle output state}, do these results define diagnostic tools for the experimental certification of the indistinguishable preparation of the injected photons.

\section{Correlation Spectroscopy}
Next, we exploit the structure of (\ref{eq:spectro}) to establish how the C-dataset can be used to probe the temporal structure of the {\em many}-particle 
input state by manipulating a {\em single} photon's input state: Assume that the injection times of the first $n-1$ photons be fixed, and that the $n$th 
photon's injection time be controllable by an adjustable delay line. Then, by virtue of the sum of exponentials in (\ref{eq:spectro}) (and, likewise, in the 
corresponding expressions (\ref{eq:PDTrainSecondMomentOri}-\ref{eq:DSecMom}, \ref{eq:final1}) for $CV$), whenever $t_n\simeq t_k$, $n\ne k$, the thus triggered two-photon interference between photons $n$ and $k$ 
will induce a dip in $NM$, of width $(\Delta\omega) ^ {-1}$, centred around $t_k$, much as in a typical spectroscopic experiment. This protocol thus even allows
the inference (with finite resolution controlled by the photons' spectral bandwidth) of the actual timing of the injected photons. 

\begin{figure}
	\includegraphics[width=0.48\textwidth]{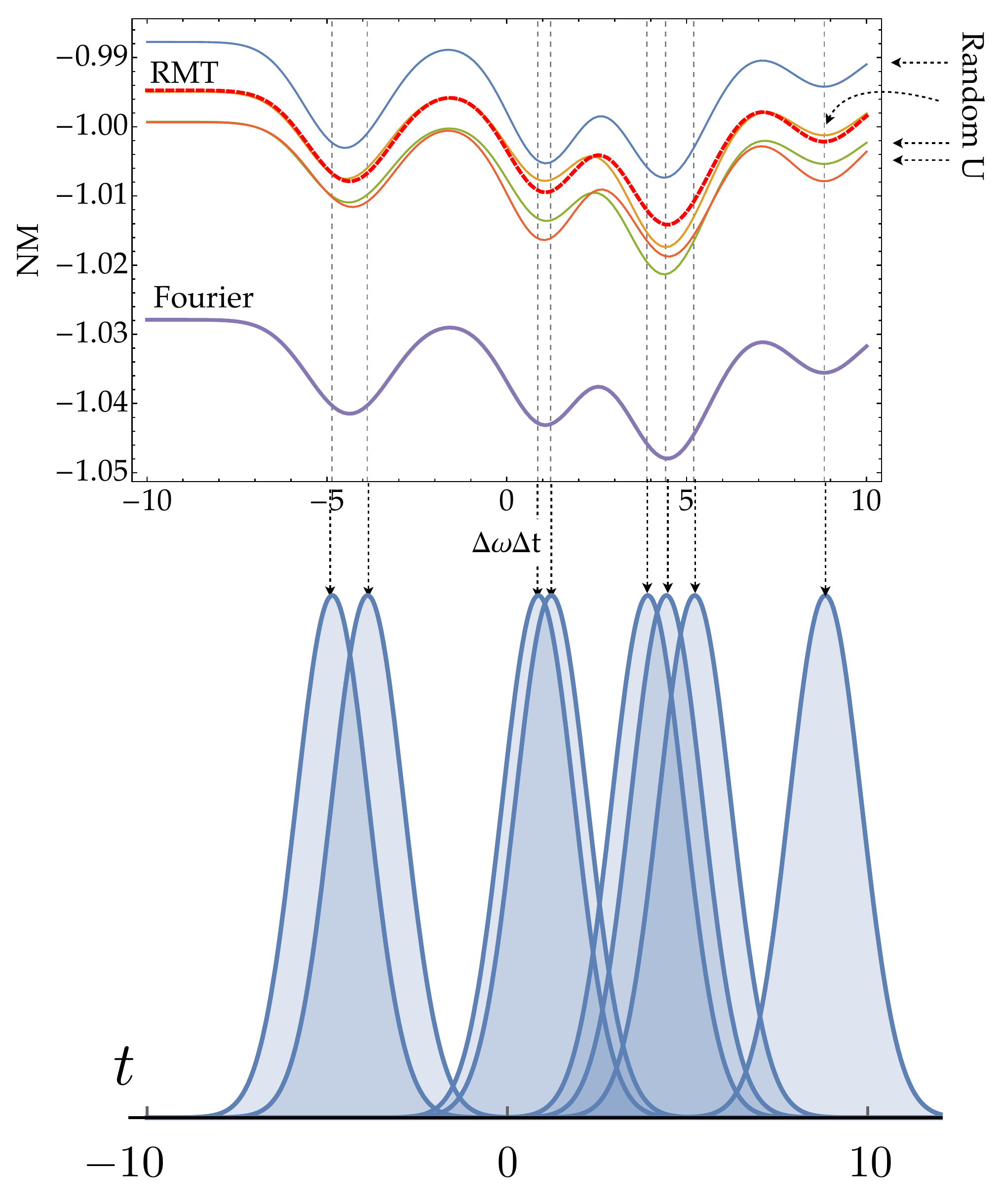}
  \caption{Normalised mean $NM$, as a function of the probe photon's (see Fig.~\ref{fig:Sketch}) delay with respect to $t=0$, and for 
  $n-1=8$ injected photons with injection times $t_k = -4.86071$, $-3.87957$, $0.858186$, $1.21835$, $3.89386$, $4.41308$, $5.19717$, $8.82249$ 
  (in units of $1/\Delta \omega$), for numerically generated, random (thin lines) $m=30$ mode unitaries, compared to the RMT prediction 
  (\ref{eq:spectro}) (thick, dashed red line), and to the result (\ref{eq:Fourier}) for the Fourier circuit. Clearly, whenever the probe photon's delay 
  coincides with any one of the other photons' injection times (associated wave packets are displayed in the bottom panel), a resonance-like 
  dip emerges in the C-dataset's lowest order statistical moment.} 
   \label{fig:Spectro}
\end{figure}

Again, as shown in Fig.~\ref{fig:Spectro}, RMT results and numerical simulations agree qualitatively very well, with some quantitative 
deviations in the vicinity of the minima of $NM$. We attribute these to the difference between the RMT average (\ref{eq:spectro})
and the contribution of the ``correlation resonance" between the probe and the $k$th input photon to the signal as generated by 
a specific realisation of $U$.

We finally stress that the statistical characterisation of the quantum-classical transition as here proposed can also be applied to structured or highly symmetric 
circuits such as described by Fourier matrices. This type of circuits exhibit prominent interference effects which have been experimentally 
demonstrated very recently \cite{crespi_suppression_2016} and can be understood analytically \cite{tichy_zero-transmission_2010,tichy_many-particle_2012,tichy_stringent_2014}. Also the $C_{ij}$ and, subsequently, 
$NM$ can be directly evaluated, {\em without}
recourse to RMT.
$NM$ of the Fourier circuit C-dataset is obtained via a direct evaluation of (\ref{eq:M1}), with $U$ a Fourier matrix, hence \begin{equation}
U_{q_k i} = \frac{1}{\sqrt{m}}\exp\left(2 \pi {\rm i} \frac{(q_k -1)(i-1)}{m}\right).
\end{equation}
We may now write (6) as
\begin{equation}\begin{split}
C_{ij} = -\frac{n}{m^2} - \frac{1}{m^2} \sum^n_{k\neq l=1} \abs{\braket{\psi_k}{\psi_l}}^2 \exp\left(2 \pi {\rm i} \frac{(q_l - q_k)(j - i)}{m}\right),
\end{split}
\end{equation}
which needs to be averaged over all output modes $i$ and $j$ to obtain $M_1$ (\ref{eq:M1}). When we consider a fixed value $i$, we obtain that
\begin{equation}\label{eq:theIdentity}
\sum_{\substack{j=1\\ j\neq i}}^m\exp\left(2 \pi {\rm i} \frac{(q_l - q_k)j}{m}\right) = - \exp\left(2 \pi {\rm i} \frac{(q_l - q_k) i}{m}\right).
\end{equation}
The identity (\ref{eq:theIdentity}) implies
\begin{equation}
\sum_{\substack{i,j = 1\\ i\neq j}}^m\exp\left(2 \pi {\rm i} \frac{(q_l - q_k)(j - i)}{m}\right) = - m,
\end{equation}
and, hence, with 
(\ref{NM},\ref{eq:M1}), one obtains
\begin{equation}\label{eq:Fourier}
NM = -1-\frac{1}{n(m-1)}\sum_{k\neq l =1}^n \exp\bigg(-\frac{{\Delta\omega}^2 (t_k - t_l)^2}{2}\bigg),
\end{equation}
with a slightly increased visibility of the signal displayed in Fig.~\ref{fig:Spectro}, as compared to the result for a random 
scatterer. 

Note that this result nicely illustrates two rather complementary aspects of multiparticle, multimode interference in the presence of symmetries: On the one hand, the Fourier circuit's symmetries induce the suppression of specific, well-defined output events, which define a highly {\em sensitive} probe of the precise implementation of the Fourier map and of the concomitant multiparticle interference. On the other hand, {\em irrespective} of these isolated output events specific to the Fourier map, there are {\em robust} statistical features proper to {\em all} output events which, both, highly symmetric and fully random circuits, have in common! However, even on the level of these statistical quantifiers does the difference between the underlying unitaries emerge, through an essentially constant shift, as evident from Fig.~\ref{fig:Spectro}. The precise connection between the specific structure of the unitaries and the observed bias remains to be elucidated.

\section{Conclusions}

Let us conclude with the observation that the treatment of many-particle interference phenomena in terms of a set of correlators with 
statistical properties which are indicative of structural properties of the injected quantum states defines a new type of {\em correlation}
spectroscopy. While we focussed here on the photonic context which originally motivated this work, the underlying theoretical structure
as incarnated by (\ref{eq:Corr}) is rather general and lends itself to straightforward generalisations to other ``distinguishing'' degrees of freedom, as well
as to other, e.g. fermionic particle species. Since the overlaps $\abs{\braket{\psi_k}{\psi_l}}^2$ in the distinguishing degree of freedom define some sort 
of which-way information on the level of two-particle transition amplitudes, this furthermore indicates new directions for the decoherence theory of quantum 
systems of indistinguishable particles.\\

{\bf Acknowledgements:} The authors thank Juan Diego Urbina and Klaus Richter for fruitful discussions. M.W. expresses gratitude to the German National Academic Foundation for financial support. A.B. acknowledges support by the EU Collaborative project QuProCS (Grant Agreement 641277). 

\bibliographystyle{apsrev}
\bibliography{PaperSpectro}

\end{document}